\documentstyle[twoside]{article}

\catcode`\@=11
\long\def\@makefntext#1{
\protect\noindent \hbox to 3.2pt {\hskip-.9pt  
$^{{\eightrm\@thefnmark}}$\hfil}#1\hfill}		%CAN BE USED 

\def\@makefnmark{\hbox to 0pt{$^{\@thefnmark}$\hss}}	%ORIGINAL 
	
\def\ps@myheadings{\let\@mkboth\@gobbletwo
\def\@oddhead{\hbox{}
\rightmark\hfil\eightrm\thepage}   
\def\@oddfoot{}\def\@evenhead{\eightrm\thepage\hfil
\leftmark\hbox{}}\def\@evenfoot{}
\def\sectionmark##1{}\def\subsectionmark##1{}}

\oddsidemargin=\evensidemargin
\newcounter{sectionc}\newcounter{subsectionc}\newcounter{subsubsectionc}
\renewcommand{\section}[1] {\vspace{12pt}\addtocounter{sectionc}{1} 
\setcounter{subsectionc}{0}\setcounter{subsubsectionc}{0}\noindent 
	{\tenbf\thesectionc. #1}\par\vspace{5pt}}
\renewcommand{\subsection}[1] {\vspace{12pt}\addtocounter{subsectionc}{1} 
	\setcounter{subsubsectionc}{0}\noindent 
	{\bf\thesectionc.\thesubsectionc. {\kern1pt \bfit #1}}\par\vspace{5pt}}
\renewcommand{\subsubsection}[1] {\vspace{12pt}\addtocounter{subsubsectionc}{1}
	\noindent{\tenrm\thesectionc.\thesubsectionc.\thesubsubsectionc.
	{\kern1pt \tenit #1}}\par\vspace{5pt}}
\newcommand{\nonumsection}[1] {\vspace{12pt}\noindent{\tenbf #1}
	\par\vspace{5pt}}

\newcounter{appendixc}
\newcounter{subappendixc}[appendixc]
\newcounter{subsubappendixc}[subappendixc]
\renewcommand{\thesubappendixc}{\Alph{appendixc}.\arabic{subappendixc}}
\renewcommand{\thesubsubappendixc}
	{\Alph{appendixc}.\arabic{subappendixc}.\arabic{subsubappendixc}}

\renewcommand{\appendix}[1] {\vspace{12pt}
        \refstepcounter{appendixc}
        \setcounter{figure}{0}
        \setcounter{table}{0}
        \setcounter{lemma}{0}
        \setcounter{theorem}{0}
        \setcounter{corollary}{0}
        \setcounter{definition}{0}
        \setcounter{equation}{0}
        \renewcommand{\thefigure}{\Alph{appendixc}.\arabic{figure}}
        \renewcommand{\thetable}{\Alph{appendixc}.\arabic{table}}
        \renewcommand{\theappendixc}{\Alph{appendixc}}
        \renewcommand{\thelemma}{\Alph{appendixc}.\arabic{lemma}}
        \renewcommand{\thetheorem}{\Alph{appendixc}.\arabic{theorem}}
        \renewcommand{\thedefinition}{\Alph{appendixc}.\arabic{definition}}
        \renewcommand{\thecorollary}{\Alph{appendixc}.\arabic{corollary}}
        \renewcommand{\theequation}{\Alph{appendixc}.\arabic{equation}}
        \noindent{\tenbf Appendix \theappendixc #1}\par\vspace{5pt}}
\newcommand{\subappendix}[1] {\vspace{12pt}
        \refstepcounter{subappendixc}
        \noindent{\bf Appendix \thesubappendixc. {\kern1pt \bfit #1}}
	\par\vspace{5pt}}
\newcommand{\subsubappendix}[1] {\vspace{12pt}
        \refstepcounter{subsubappendixc}
        \noindent{\rm Appendix \thesubsubappendixc. {\kern1pt \tenit #1}}
	\par\vspace{5pt}}

\newcommand{\textlineskip}{\baselineskip=13pt}
\newcommand{\smalllineskip}{\baselineskip=10pt}

\def\eightcirc{
\begin{picture}(0,0)
\put(4.4,1.8){\circle{6.5}}
\end{picture}}
\def\eightcopyright{\eightcirc\kern2.7pt\hbox{\eightrm c}}

\def\abstracts#1#2#3{{
	\centering{\begin{minipage}{4.5in}\baselineskip=10pt\footnotesize
	\parindent=0pt #1\par 
	\parindent=15pt #2\par
	\parindent=15pt #3
	\end{minipage}}\par}} 

\renewenvironment{thebibliography}[1]
	{\frenchspacing
	 \ninerm\baselineskip=11pt
	 \begin{list}{\arabic{enumi}.}
	{\usecounter{enumi}\setlength{\parsep}{0pt}
	 \setlength{\leftmargin 12.7pt}{\rightmargin 0pt} %FOR 1--9 ITEMS
	 \setlength{\itemsep}{0pt} \settowidth
	{\labelwidth}{#1.}\sloppy}}{\end{list}}

\newcounter{itemlistc}
\newcounter{romanlistc}
\newcounter{alphlistc}
\newcounter{arabiclistc}

\newcommand{\fcaption}[1]{
        \refstepcounter{figure}
        \setbox\@tempboxa = \hbox{\footnotesize Fig.~\thefigure. #1}
        \ifdim \wd\@tempboxa > 5in
           {\begin{center}
        \parbox{5in}{\footnotesize\smalllineskip Fig.~\thefigure. #1}
            \end{center}}
        \else
             {\begin{center}
             {\footnotesize Fig.~\thefigure. #1}
              \end{center}}
        \fi}

\newcommand{\tcaption}[1]{
        \refstepcounter{table}
        \setbox\@tempboxa = \hbox{\footnotesize Table~\thetable. #1}
        \ifdim \wd\@tempboxa > 5in
           {\begin{center}
        \parbox{5in}{\footnotesize\smalllineskip Table~\thetable. #1}
            \end{center}}
        \else
             {\begin{center}
             {\footnotesize Table~\thetable. #1}
              \end{center}}
        \fi}

\def\@citex[#1]#2{\if@filesw\immediate\write\@auxout
	{\string\citation{#2}}\fi
\def\@citea{}\@cite{\@for\@citeb:=#2\do
	{\@citea\def\@citea{,}\@ifundefined
	{b@\@citeb}{{\bf ?}\@warning
	{Citation `\@citeb' on page \thepage \space undefined}}
	{\csname b@\@citeb\endcsname}}}{#1}}

\newif\if@cghi
\def\cite{\@cghitrue\@ifnextchar [{\@tempswatrue
	\@citex}{\@tempswafalse\@citex[]}}
\def\citelow{\@cghifalse\@ifnextchar [{\@tempswatrue
	\@citex}{\@tempswafalse\@citex[]}}
\def\@cite#1#2{{$\null^{#1}$\if@tempswa\typeout
	{IJCGA warning: optional citation argument 
	ignored: `#2'} \fi}}

\def\pmb#1{\setbox0=\hbox{#1}
	\kern-.025em\copy0\kern-\wd0
	\kern.05em\copy0\kern-\wd0
	\kern-.025em\raise.0433em\box0}

\def\fnt#1#2{\footnotetext{\kern-.3em
	{$^{\mbox{\scriptsize #1}}$}{#2}}}

\font\tenrm=cmr10
\font\tenit=cmti10 
\font\tenbf=cmbx10
\font\bfit=cmbxti10 at 10pt
\font\ninerm=cmr9

\font\eightrm=cmr8

\def\qed{\hbox{${\vcenter{\vbox{			%HOLLOW SQUARE
   \hrule height 0.4pt\hbox{\vrule width 0.4pt height 6pt
   \kern5pt\vrule width 0.4pt}\hrule height 0.4pt}}}$}}

\begin{document}

%\runninghead{$\ldots$} {$\ldots$}

\normalsize\textlineskip
\setcounter{page}{1}

%\copyrightheading{}			%{Vol. 0, No. 0 (1993) 000--000}

%\vspace*{0.88truein}

%\fpage{}
\centerline{\bf THE ROLE OF DIAGONALIZATION}
\vspace*{0.035truein}
\centerline{\bf WITHIN A DIAGONALIZATION/MONTE CARLO SCHEME \footnote{To appear in the proceedings of               
DPF2000, Columbus, August 2000.}%
}

\vspace*{0.37truein}
\centerline{\footnotesize Dean Lee}
\vspace*{0.015truein}
\centerline{\footnotesize\it Department of Physics, Univ. of Massachusetts, Amherst, MA 01003}

\vspace*{0.21truein}
\abstracts{We discuss a method called quasi-sparse eigenvector diagonalization which finds the most important basis vectors of the low energy eigenstates of a
quantum Hamiltonian.  It can operate using any basis, either orthogonal or
non-orthogonal, and any sparse Hamiltonian, either Hermitian, non-Hermitian,
finite-dimensional, or infinite-dimensional.  The method is part of a new
computational approach which combines both diagonalization and Monte Carlo
techniques.}{}{}

\vspace*{1pt}\textlineskip
\section{Introduction}
\noindent
The methods of Monte Carlo and diagonalization are almost opposite in their
strengths and weaknesses. \ Monte Carlo requires relatively little storage,
can be performed using parallel processors, and in some cases the
computational effort scales reasonably with system size. \ But it has great
difficulty for systems with sign or phase oscillations and provides only
indirect information on wavefunctions and excited states. \ Diagonalization
methods on the other hand do not suffer from fermion sign problems, can handle
complex-valued actions, and can extract details of the spectrum and eigenstate
wavefunctions. \ However the required memory and CPU time scales exponentially
with the size of the system.

Recently we proposed a new computational approach which takes advantage of the strengths of the two methods in their respective domains.\cite{qse,sec}
\ The first half of the approach involves finding and diagonalizing the
Hamiltonian restricted to an optimal subspace. \ This subspace is designed to
include the most important basis vectors of the lowest energy eigenstates.
\ Once the most important basis vectors are found and their interactions
treated exactly, Monte Carlo is used to sample the contribution of the
remaining basis vectors. \ Much of the sign problem is negated by treating the
interactions of the most important basis states exactly, while storage and CPU
problems are resolved by stochastically sampling the collective effect of the
remaining states.

In our approach diagonalization is used as the starting point of the Monte
Carlo calculation. \ Unfortunately we find that most diagonalization methods
are either not sufficiently general, not able to search an infinite or large
dimensional Hilbert space, not efficient at finding important basis vectors,
or not compatible with the subsequent Monte Carlo part of the calculation.
\ In this brief article we review a new diagonalization method called
quasi-sparse eigenvector (QSE) diagonalization. \ It is a general algorithm
which can operate using any basis, either orthogonal or non-orthogonal, and
any sparse Hamiltonian, either real, complex, Hermitian, non-Hermitian,
finite-dimensional, or infinite-dimensional. \ It is able to find the most
important basis states of several low energy eigenvectors simultaneously,
including those with identical quantum numbers, from a random start with no
prior knowledge about the form of the eigenvectors.

\section{Quasi-sparse eigenvector method}
\noindent
Let $\left|  e_{i}\right\rangle $ denote a complete set of basis vectors.
\ For a given energy eigenstate
\begin{equation}
|v\rangle=\sum_{i}c_{i}\left|  e_{i}\right\rangle ,
\end{equation}
the important basis states of $|v\rangle$ are defined to be those $\left|
e_{i}\right\rangle $ such that for fixed normalizations of $|v\rangle$ and the
basis states, $\left|  c_{i}\right|  $ exceeds a prescribed threshold value.
\ If $|v\rangle$ can be well-approximated by the contribution from only its
important basis states we refer to the eigenvector $|v\rangle$ as
\textit{quasi-sparse} with respect to $\left|  e_{i}\right\rangle $.

Standard sparse matrix algorithms such as the Lanczos or Arnoldi methods allow
one to find the extreme eigenvalues and eigenvectors of a sparse matrix
efficiently, without having to store or manipulate large non-sparse matrices.
\ However in quantum field theory or many body theory one considers very large
or infinite dimensional spaces where even storing the components of a general
vector is impossible. \ For these more difficult problems the strategy is to
approximate the low energy eigenvectors of the large space by diagonalizing
smaller subspaces. \ If one has sufficient intuition about the low energy
eigenstates it may be possible to find a useful truncation of the full vector
space to an appropriate smaller subspace. \ In most cases, however, not enough
is known \textit{a priori }about the low energy eigenvectors. \ The dilemma is
that to find the low energy eigenstates one must truncate the vector space,
but in order to truncate the space something must be known about the low
energy states.

Our solution to this puzzle is to find the low energy eigenstates and the
appropriate subspace truncation at the same time by a recursive process. \ We
call the method quasi-sparse eigenvector (QSE) diagonalization, and we
describe the steps of the algorithm as follows. \ The starting point is any
complete basis for which the Hamiltonian matrix $H_{ij}$ is sparse. \ The
basis vectors may be non-orthogonal and/or the Hamiltonian matrix may be
non-Hermitian. \ The following steps are iterated:

\begin{enumerate}
\item  Select a subset of basis vectors $\left\{  e_{i_{1}},\cdots,e_{i_{n}%
}\right\}  $ and call the corresponding subspace $S$.

\item  Diagonalize $H$ restricted to $S$ and find one eigenvector $v$.

\item  Sort the basis components of $v$ according to their magnitude and
remove the least important basis vectors.

\item  Replace the discarded basis vectors by new basis vectors. \ These are
selected at random according to some weighting function from a pool of
candidate basis vectors which are connected to the old basis vectors through
non-vanishing matrix elements of $H$.

\item  Redefine $S$ as the subspace spanned by the updated set of basis
vectors and repeat steps 2 through 5.
\end{enumerate}

If the subset of basis vectors is sufficiently large, then it can be shown
that the exact low energy eigenvectors are stable fixed points of the QSE
update process.\cite{qse} \ As the name indicates the accuracy of the
quasi-sparse eigenvector method depends on the quasi-sparsity of the low
energy eigenstates in the chosen basis. \ If the eigenvectors are quasi-sparse
then the QSE method provides an efficient way to find the important basis
vectors. \ In the context of our diagonalization/Monte Carlo approach, this
means that diagonalization does most of the work and only a small amount of
correction is needed. \ This correction is found by Monte Carlo sampling the
remaining basis vectors, a technique called stochastic error correction.\cite{sec}  If however the eigenvectors are not quasi-sparse then one must
rely more heavily on the Monte Carlo portion of the calculation. \ The fastest
and most reliable way we know to determine whether the low energy eigenstates
of a Hamiltonian are quasi-sparse with respect to a chosen basis is to use the
QSE algorithm and look at the results of the successive iterations. We generally find that eigenvectors are quasi-sparse
with respect to a chosen basis if the spacing between energy levels is not too
small compared with the size of the off-diagonal entries of the Hamiltonian matrix.\cite{qse}

We regard QSE diagonalization as only a starting point for the Monte Carlo
part of the calculation. \ Once the most important basis vectors are found and
their interactions treated exactly, a technique called stochastic error
correction is used to sample the contribution of the remaining basis vectors.\cite{sec}.

\nonumsection{Acknowledgements}
\noindent
The author thanks all collaborators on the works cited here and the organizers
and participants of the DPF2000 meeting in Columbus.  Financial support
provided by the National Science Foundation.

\nonumsection{References}


\begin{thebibliography}{9}
\bibitem{qse}D. J. Lee, N. Salwen, D. D. Lee, hep-th/0002251.

\bibitem{sec}D. Lee, N. Salwen, M. Windoloski, hep-lat/0010039.

\bibitem{tutorial}D. Lee, hep-th/0007010.\bigskip
\end{thebibliography}
\end{document}